\documentclass{aip-cp}

\usepackage[numbers]{natbib}
\usepackage{rotating}
\usepackage{graphicx}

\newcommand{\gev}{\rm GeV}

\newcommand{\mev}{\rm MeV}
\newcommand{\mevcs}{{\rm MeV}/c^2}

\newcommand{\xyz}{\rm XYZ}

\newcommand{\y}{Y(4260)}

\newcommand{\zc}{Z_c(3900)}

\newcommand{\zb}{Z_b(10610)}
\newcommand{\zbp}{Z_b(10650)}

\newcommand{\BR}{{\cal B}}

\newcommand{\psp}{\psi(2S)}

\newcommand{\jpsi}{J/\psi}

\newcommand{\EE}{e^+e^-}
\newcommand{\MM}{\mu^+\mu^-}
\newcommand{\LL}{\ell^+\ell^-}

\newcommand{\pp}{\pi^+\pi^-}

\newcommand{\kk}{K^+K^-}

\newcommand{\kkjpsi}{K^+K^- J/\psi}
\newcommand{\ppjpsi}{\pi^+\pi^- J/\psi}

\newcommand{\pppsp}{\pi^+\pi^- \psp}

\newcommand{\ppns}{\pi^+\pi^- \Upsilon(nS)}
\newcommand{\ppones}{\pi^+\pi^- \Upsilon(1S)}
\newcommand{\pptwos}{\pi^+\pi^- \Upsilon(2S)}
\newcommand{\ppthrees}{\pi^+\pi^- \Upsilon(3S)}
\newcommand{\ppnp}{\pi^+\pi^- h_b(nP)}
\newcommand{\pponep}{\pi^+\pi^- h_b(1P)}
\newcommand{\pptwop}{\pi^+\pi^- h_b(2P)}
\newcommand{\bbpi}{B^{(*)}\bar{B^{(*)}}\pi}
\newcommand{\beq}{\begin{equation}}
\newcommand{\eeq}{\end{equation}}
\newcommand{\bitm}{\begin{itemize}}
\newcommand{\eitm}{\end{itemize}}
\begin{document}

%%\preprint{INT-PUB-15-072}
% Title portion
\title{New results on the $\xyz$ states from Belle experiment}

\author[aff1]{Chang-Zheng Yuan\corref{cor1} for
the Belle Collaboration}

\affil[aff1]{Institute of High Energy Physics, Chinese Academy of
Sciences, Beijing 100049, China} \corresp[cor1]{Corresponding
author: yuancz@ihep.ac.cn}

\maketitle

\begin{abstract}
We review the results on the $\xyz$ states from the Belle
experiment, including the measurement of inclusive hadronic cross
sections and exclusive cross sections in $\EE$ annihilation and
the $Z_b$ states in the bottomonium energy region; and the study
of the $Y$ states and the $Z_c$ states in charmonium energy
region.
\end{abstract}

\section{INTRODUCTION}

Many charmonium and charmonium-like states were discovered at
$B$-factories in the first decade of the 21st century followed by
similar discoveries in the bottomonium and bottomniumlike
states~\cite{PBFB}. Whereas some of these are good quarkonium
candidates, as predicted in different models, many states have
exotic properties, which may indicate that exotic states, such as
multi-quark, molecule, hybrid, or hadron-quarkonium, have been
observed~\cite{review}. In such studies, the Belle
experiment~\cite{Belle} at the KEKB asymmetric-energy $e^+e^-$
collider ($3.5~\gev$ $e^+$ and $8.0~\gev$ $e^-$)~\cite{KEKB}
played a leading role.

The Belle detector is a large-solid-angle magnetic spectrometer
that consists of a silicon vertex detector, a 50-layer central
drift chamber, an array of aerogel threshold Cherenkov counters, a
barrel-like arrangement of time-of-light scintillation counters,
and an electromagnetic calorimeter comprised of CsI(Tl) crystals
located inside a super-conducting solenoid coil that provides a
1.5T magnetic field. An iron flux return located outside of the
coil is instrumented to detect $K^0_{\rm L}$ mesons and to
identify muons.

Belle experiment accumulated 1014~fb$^{-1}$ data at different
$\Upsilon$ peaks and at the continuum energies nearby. The main
data samples are summarized in Table~\ref{datasample}.

\begin{table}[htb]
\caption{Data samples collected at Belle experiment.}
 \label{datasample}
\begin{tabular}{rrr} \hline
 $\sqrt{s}$~(GeV) & ${\cal L}$ (fb$^{-1}$) & description of the sample  \\
  \hline
  9.460  &    6~~~    &  $(102\pm 2)\times 10^6$ $\Upsilon(1S)$ events\\
 10.023  &   25~~~    &  $(158\pm 4)\times 10^6$ $\Upsilon(2S)$ events\\
 10.355  &    3~~~    &  $(11.0\pm 0.3)\times 10^6$ $\Upsilon(3S)$ events\\
 10.580  &  702~~~    &  $(772\pm 10)\times 10^6$ $\Upsilon(4S)$ events\\
 10.867  &  121~~~    &  $(7.1\pm 1.3)\times 10^6$ $\Upsilon(5S)$ events\\
 10.520  &   89~~~    &  off resonance \\
 10.63-11.05  &   28~~~    &  $\Upsilon(5S)$ scan at 83 energies \\ \hline
\end{tabular}
\end{table}

\section{BOTTOMONIUM and BOTTOMONIUMLIKE STATES}

With the 83 scan data points for $\sqrt{s}=10.63$--$11.05$~GeV,
and the larger data sample at the $\Upsilon(5S)$ peak, we measured
the inclusive cross section of $\EE\to b\bar{b}$~\cite{rb}, the
exclusive cross sections of $\EE\to \ppns$~\cite{rb},
$\ppnp$~\cite{hbpipi}, clear resonant structures are observed
which correspond to the $5S$ and $6S$ bottomonium states. The
intermediate states in the three-body decays of $\Upsilon(5S)\to
\ppns$~\cite{zb_pwa}, $\ppnp$, as well as $\bbpi$ are studied,
information on charged bottomonium states, $\zb$ and $\zbp$, are
obtained.

The data consist of 121.4~fb$^{-1}$ at $\sqrt{s}=10.865$~GeV;
approximately 1~fb$^{-1}$ at each of the 22 energy points between
10.63 and 11.02~GeV; and 50~pb$^{-1}$ at each of 61 points taken
in 5~MeV steps between 10.75 and 11.05~GeV. The non-resonant
$q\bar q$ continuum $(q\in\{u,d,s,c\})$ background is obtained
using a 1.03~fb$^{-1}$ data sample taken at $\sqrt{s}=10.52$~GeV.

\subsection{\boldmath $R_b$ Measurement}

\begin{figure}[htb]
 \centering
 \includegraphics[height=8cm]{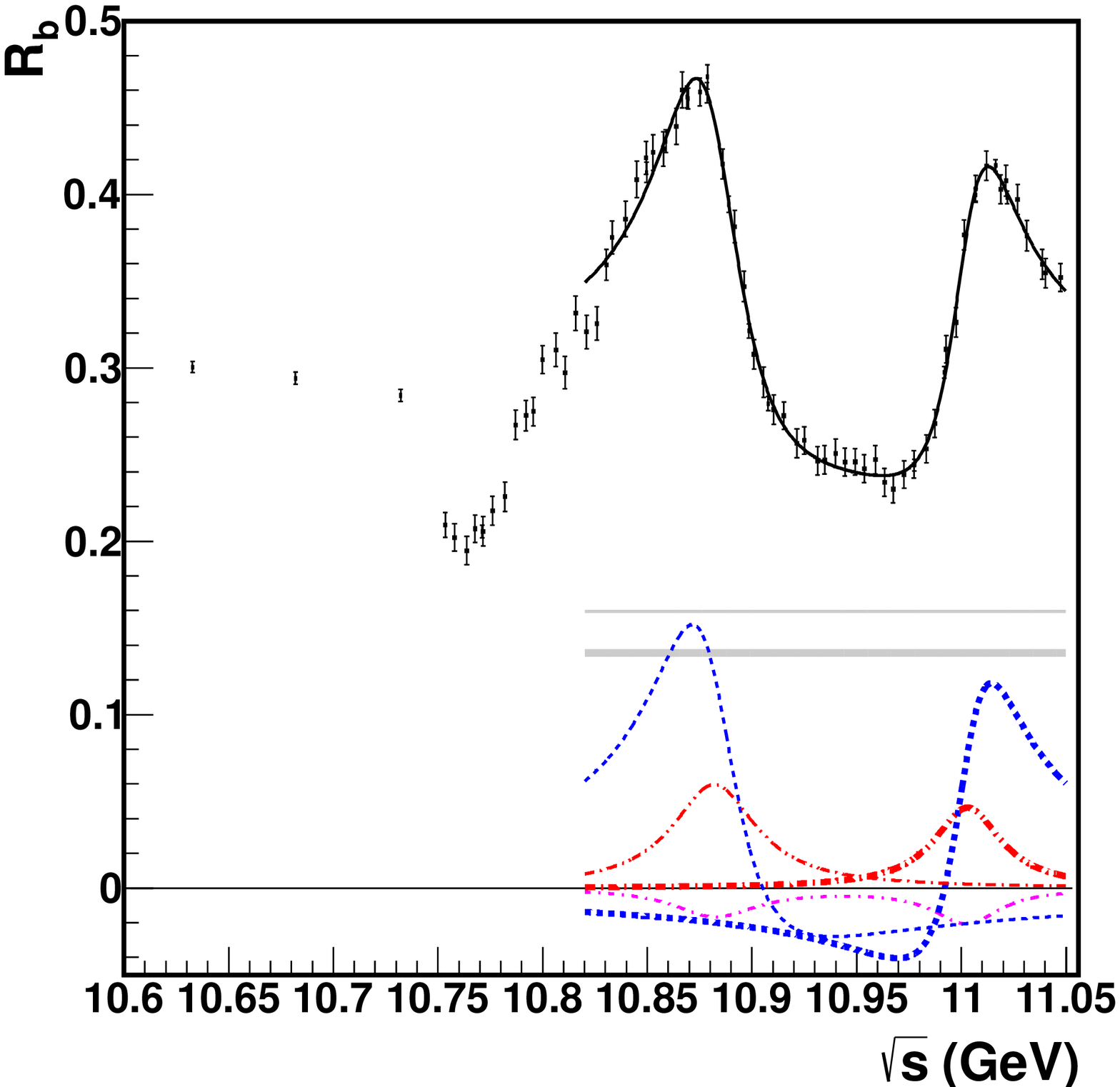}
 \includegraphics[height=8cm]{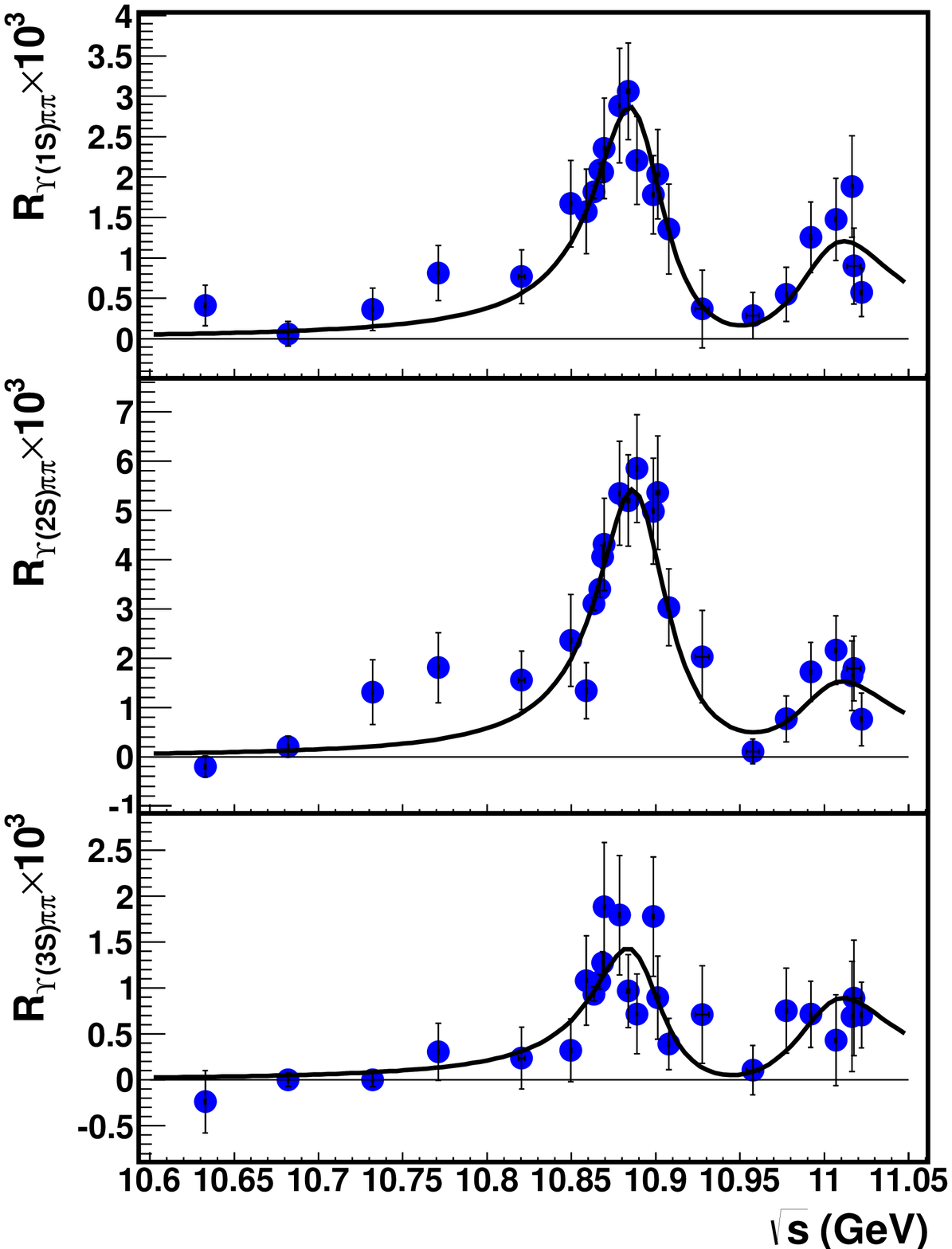}
\caption{$R_b$ data with components of fit (left) and $\EE\to
\ppns$ data (right) for $\Upsilon(1S)$ (top), $\Upsilon(2S)$
(center), and $\Upsilon(3S)$ (bottom), and the fit with coherent
sum of two Breit-Wigner functions. Error bars are statistical
only.} \label{fig:rb}
\end{figure}

This measurement was done with all the data samples above
10.63~GeV. The reduced $\EE\to b\bar{b}$ cross section is defined
as $R_b\equiv \frac{\sigma(\EE\to b\bar{b})}{\sigma^0(\EE\to
\MM)}$, where the denominator is the Born cross section of $\EE\to
\MM$. In this analysis~\cite{rb}, we select hadronic events and
subtract the non-$b\bar{b}$ background and initial state radiation
(ISR) produced narrow bottomonium states, then do efficiency
correction. The resulting $R_b$ and the fit with $\Upsilon(5S)$
and $\Upsilon(6S)$ are shown in Fig.~\ref{fig:rb}(left). From the
fit to $R_b$, we find $M_{10860}=(10881.8^{+1.0}_{-1.1}\pm
1.2)$~MeV/$c^2$, $\Gamma_{10860} =
(48.5^{+1.9}_{-1.8}\,^{+2.0}_{-2.8})$~MeV, $M_{11020}=(11003.0\pm
1.1^{+0.9}_{-1.0})$~MeV/$c^2$, $\Gamma_{11020} =
(39.3^{+1.7}_{-1.6}\,^{+1.3}_{-2.4})$~MeV, and
$\phi_{11020}-\phi_{10860} = (-1.87^{+0.32}_{-0.51}\pm 0.16)$~rad.
The fit results depend strongly on the parametrization of the line
shape and the fitting range as there are many bottom meson pair
thresholds in this energy region.

\subsection{\boldmath $\EE\to \ppns$ and $Z_b\to \pi^\pm \Upsilon(nS)$}

The cross sections of $\EE\to \ppns$ ($n$ = 1, 2, 3) at
$\Upsilon(5S)$ peak and 22 energy points between 10.63 and
11.02~GeV with approximately 1~fb$^{-1}$ each are measured.

Candidate $\Upsilon(nS)[\to \mu^+\mu^-]\pi^+\pi^-$ events are
selected for the measurement. Figure~\ref{fig:rb}(right) shows the
$R_{\ppns}\equiv \sigma(e^+e^-\to \ppns)/\sigma^0(\EE\to \MM)$.
The cross sections are fit for the masses and widths of the
$\Upsilon(5S)$ and $\Upsilon(6S)$ resonances. Unlike $R_b$, which
includes a large non-resonant $b\bar{b}$ component, it is found
that $\EE\to \ppns$ is dominated by the two resonances. With
$\EE\to \ppns$, we measure $M_{10860}=(10891.1\pm
3.2^{+0.6}_{-1.5})$~MeV/$c^2$ and $\Gamma_{10860} =
(53.7^{+7.1}_{-5.6}\,^{+0.9}_{-5.4})$~MeV, and report the first
measurements
$M_{11020}=(10987.5^{+6.4}_{-2.5}\,^{+9.0}_{-2.1})$~MeV/$c^2$,
$\Gamma_{11020} = (61^{+9}_{-19}\,^{+2}_{-20})$~MeV, and the
relative phase $\phi_{\rm 11020}-\phi_{\rm 10860} = (-1.0\pm
0.4\,^{+1.0}_{-0.1})$ rad.

The large statistics at the $\Upsilon(5S)$ peak make a study of
the intermediate states of $\EE\to \ppns$ possible~\cite{zb_pwa}.
After all the selections, we are left with 1905, 2312, and 635
candidate events for the $\ppones$, $\pptwos$, and $\ppthrees$
final states, respectively. We performed a full amplitude analysis
of three-body $\EE\to \ppns$ transitions and determined the
relative fractions of various quasi-two-body components of the
three-body amplitudes as well as the spin and parity of the two
observed $Z_b$ states. The favored quantum numbers are $J^P=1^+$
for both $\zb$ and $\zbp$ states while the alternative $J^P=1^-$
and $J^P=2^\pm$ combinations are rejected at confidence levels
exceeding 6 standard deviations.

Results of the amplitude analysis are summarized in
Table~\ref{tab:results}, where fractions of individual
quasi-two-body modes, masses and widths of the two $Z_b$ states,
the relative phase, $\phi_Z$, between the two $Z_b$ amplitudes and
fraction $c_{Z_{10610}}/c_{Z_{10650}}$ of their amplitudes are
given.

\begin{table*}[htb]
\caption{Summary of results of fits to $\ppns$ events in the
signal regions.} \label{tab:results}
  \centering
  \begin{tabular}{lccc} \hline
Parameter & ~~~~$\ppones$~~~~   &
              ~~~~$\pptwos$~~~~   &
              ~~~~$\ppthrees$~~~~
\\ \hline
%    $f_{Z^\mp_b(10610)\pi^\pm}{\cal{B}}(Z^\mp_b(10610)\to\Un\pi^\mp)$, \%  &
           $f_{Z^\mp_b(10610)\pi^\pm}$, \%  &
           $4.8\pm1.2^{+1.5}_{-0.3}$   &
           $18.1\pm3.1^{+4.2}_{-0.3}$   &
           $30.0\pm6.3^{+5.4}_{-7.1}$
\\
           $Z_b(10610)$ mass, MeV/$c^2$  &
         ~~$10608.5\pm3.4^{+3.7}_{-1.4}$~~     &
         ~~$10608.1\pm1.2^{+1.5}_{-0.2}$~~     &
         ~~$10607.4\pm1.5^{+0.8}_{-0.2}$~~
\\
           $Z_b(10610)$ width, MeV  &
         ~~$18.5\pm5.3^{+6.1}_{-2.3}$~~    &
         ~~$20.8\pm2.5^{+0.3}_{-2.1}$~~    &
         ~~$18.7\pm3.4^{+2.5}_{-1.3}$~~
\\
%    $f_{Z^\mp_b(10650)\pi^\pm}{\cal{B}}(Z^\mp_b(10650)\to\Un\pi^\mp)$, \%  &
           $f_{Z^\mp_b(10650)\pi^\pm}$, \%  &
           $0.87\pm0.32^{+0.16}_{-0.12}$    &
           $4.05\pm1.2^{+0.95}_{-0.15}$    &
           $13.3\pm3.6^{+2.6}_{-1.4}$
\\
           $Z_b(10650)$ mass, MeV/$c^2$  &
         ~~$10656.7\pm5.0^{+1.1}_{-3.1}$~~     &
         ~~$10650.7\pm1.5^{+0.5}_{-0.2}$~~     &
         ~~$10651.2\pm1.0^{+0.4}_{-0.3}$~~
\\
           $Z_b(10650)$ width, MeV  &
         ~~$12.1^{+11.3+2.7}_{-4.8-0.6}$~~    &
         ~~$14.2\pm3.7^{+0.9}_{-0.4}$~~    &
         ~~$ 9.3\pm2.2^{+0.3}_{-0.5}$~~
\\
           $\phi_{Z}$, degrees      &
           $67\pm36^{+24}_{-52}$     &
           $-10\pm13^{+34}_{-12}$    &
           $-5\pm22^{+15}_{-33}$
\\
           $c_{Z_b(10650)}/c_{Z_b(10610)}$   &
           $0.40\pm0.12^{+0.05}_{-0.11}$  &
           $0.53\pm0.07^{+0.32}_{-0.11}$  &
           $0.69\pm0.09^{+0.18}_{-0.07}$
\\
%           $f_{\Un f_2(1270)}{\cal{B}}(f_2(1270)\to\pp)$, \%  &
           $f_{\Upsilon(nS) f_2(1270)}$, \%  &
           $14.6\pm1.5^{+6.3}_{-0.7}$      &
           $4.09\pm1.0^{+0.33}_{-1.0}$     &
           $-$
\\
           $f_{\Upsilon(nS)(\pp)_S}$, \%  &
           $86.5\pm3.2^{+3.3}_{-4.9}$     &
           $101.0\pm4.2^{+6.5}_{-3.5}$    &
           $44.0\pm6.2^{+1.8}_{-4.3}$
\\
%           ~~~$f_{\Un f_0(980)}{\cal{B}}(f_0(980)\to\pp)$, \%  &
           ~~~$f_{\Upsilon(nS) f_0(980)}$, \%  &
           $6.9\pm1.6^{+0.8}_{-2.8}$     &
           $-$    &
           $-$
\\
\hline
\end{tabular}
\end{table*}

\subsection{\boldmath $\EE\to \ppnp$ and $Z_b\to \pi^\pm h_b(nP)$}

We measure $\EE\to \ppnp$ ($n=$1, 2) with on-resonance
$\Upsilon(5S)$ data of 121.4~fb$^{-1}$ taken in three closely
spaced energy points near 10.866~GeV, and energy scan data in the
range from about 10.77 to 11.02~GeV taken at 19 points of about
1~fb$^{-1}$ each.

The processes $\EE\to \ppnp$ are reconstructed inclusively using
the missing mass of $\pp$ pairs, $M^{\rm
miss}_{\pp}=\sqrt{(\sqrt{s}-E_{\pp}^*)^2-p_{\pp}^{*2}}$, where
$E^*_{\pp}$ and $p^*_{\pp}$ are the energy and momentum of the
$\pp$ pair measured in the center-of-mass (CM) frame.

\begin{figure}[htb]
\includegraphics[width=0.48\textwidth]{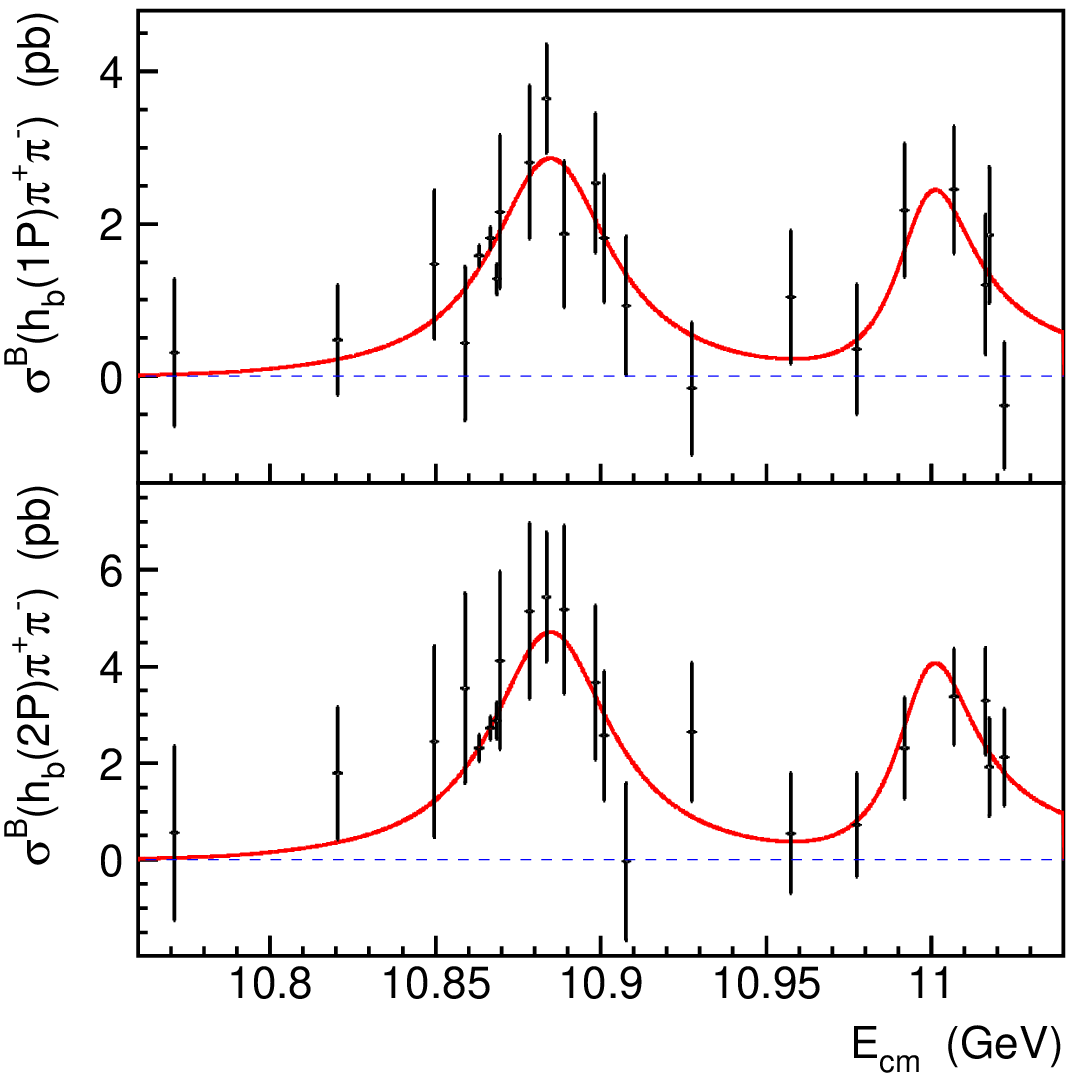}
\caption{Cross sections for the $\EE\to \pponep$ (top) and $\EE\to
\pptwop$ processes as a function of CM energy. Points with error
bars are the data, red solid curves are the fit results. }
\label{hbpp_born_simfit}
\end{figure}

The resulting cross sections are shown in
Fig.~\ref{hbpp_born_simfit}. We perform a simultaneous fit to the
energy dependence of the $\EE\to \ppnp$ $(n=1,2)$ cross sections.
The fit function is a coherent sum of two Breit-Wigner (BW)
amplitudes and (optionally) a constant with an energy continuum
contribution:
\begin{equation}
A_n\;f(s)\;|BW(s,M_5,\Gamma_5) +a\,e^{i\,\phi}BW(s,M_6,\Gamma_6)
+b\,e^{i\,\delta}|^2, \label{eq:fit_fun}
\end{equation}
where $f(s)$ is the phase space function, which is calculated
numerically taking into account the measured $Z_b$ line shape,
$BW(s,M,\Gamma)$ is a BW amplitude
$BW(s,M,\Gamma)=M\Gamma/(s-M^2+iM\Gamma)$. The parameters $A_1$,
$A_2$, $M_5$, $\Gamma_5$, $M_6$, $\Gamma_6$, $a$, $\phi$ and
(optionally) $b$, $\delta$ are floated in the fit.

We find that the significance of the non-resonant continuum
contribution is $1.5\sigma$ only. Thus the default fit function
does not include the continuum contribution. The fit results for
the default model are given in Table~\ref{tab:5_6_par}.

\begin{table}[htb]
\caption{Fit results for the default model.} \label{tab:5_6_par}
\begin{tabular}{lc}
\hline
 Parameter        & results  \\ \hline
$M_5$~(MeV/$c^2$)     & $10884.7^{+3.2}_{-2.9}{^{+8.6}_{-0.6}}$  \\
$\Gamma_5$~(MeV) & $44.2^{+11.9}_{-7.8}{^{+2.2}_{-15.8}}$  \\
$M_6$~(MeV/$c^2$)     & $10998.6\pm6.1{^{+16.1}_{-1.1}}$ \\
$\Gamma_6$~(MeV) & $29^{+20}_{-12}{^{+2}_{-7}}$  \\
$A_1 / 10^3$        & $4.8^{+2.7}_{-0.8}$   \\
$A_2 / 10^3$        & $8.0^{+4.6}_{-1.3}$   \\
$a$                 & $0.64^{+0.37}_{-0.11}{^{+0.13}_{-0}}$   \\
$(\phi/\pi)$        & $0.1^{+0.3}_{-0.5}$   \\ \hline
\end{tabular}
\end{table}

\begin{figure*}[htb]
\centering
\includegraphics[width=0.4\textwidth]{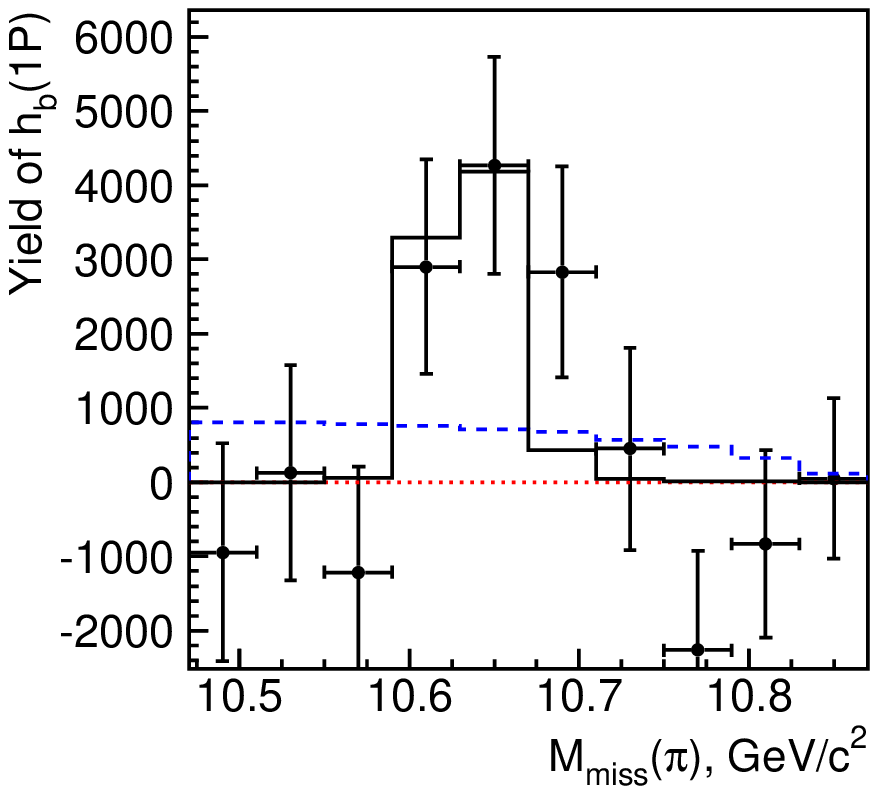}
\includegraphics[width=0.4\textwidth]{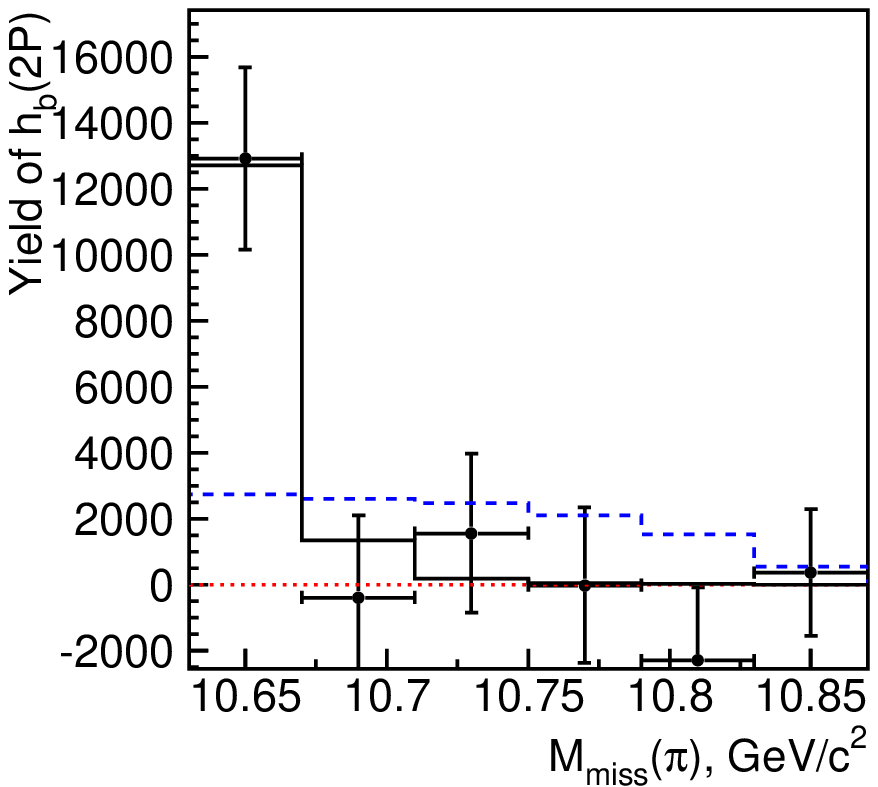}
\caption{The efficiency corrected yields of $\pp h_b(1P)$ (left)
and $\pp h_b(2P)$ (right) as a function of $pi$ missing mass.
Points represent data, the black solid histogram represents the
fit result with the shape fixed from the $\Upsilon(5S)$ analysis,
the blue dashed histogram is the result of the fit to the phase
space distribution. } \label{hb_vs_mmp_fits}
\end{figure*}

The $\pi h_b(1P)$ and $\pi h_b(2P)$ invariant mass distributions
corrected for the reconstruction efficiency are shown in
Fig.~\ref{hb_vs_mmp_fits}. The data do not follow a phase space
distribution but populate the mass region of the $\zb$ and $\zbp$
states. We find that the transitions are dominated by the
intermediate $\zb$ and $\zbp$ states, but the limited statistics
do not allow a measurement of the contribution from each mode.

%%\subsection{\boldmath $\EE\to \bbpi$ and $Z_b\to \pi^\pm B^{(*)}$}

\section{CHARMMONIUM and CHARMONIUMLIKE STATES}

The vector charmonium and charmoniumlike states are studied via
ISR at Belle in the process $e^+e^- \to \gamma_{\rm ISR}
\pi^+\pi^-J/\psi$~\cite{belley_new}, $e^+e^- \to \gamma_{\rm ISR}
\pi^+\pi^-\psi(2S)$~\cite{belle_pppsp_new}, and $e^+e^- \to
\gamma_{\rm ISR} K^+K^-J/\psi$~\cite{belle_kkjpsi_new}. Charged
charmonium states decay into a charmonium and a charged pion (or a
charged kaon) are searched for in the intermediate states.

\subsection{\boldmath $\EE\to \ppjpsi$ and $Z_c\to \pi^\pm \jpsi$}

\begin{figure}[htb]
\centering
\includegraphics[height=6cm]{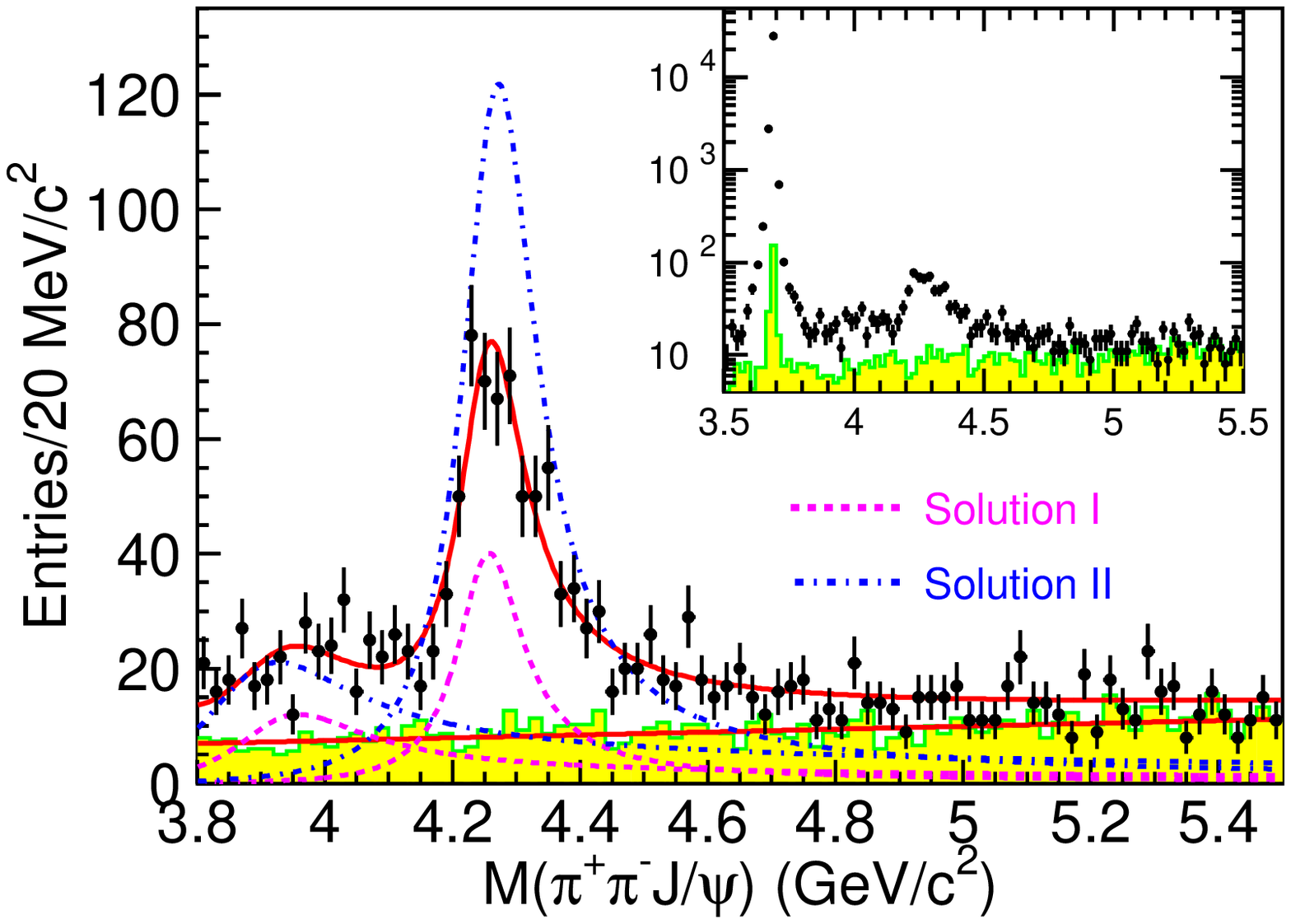}
\includegraphics[height=6cm]{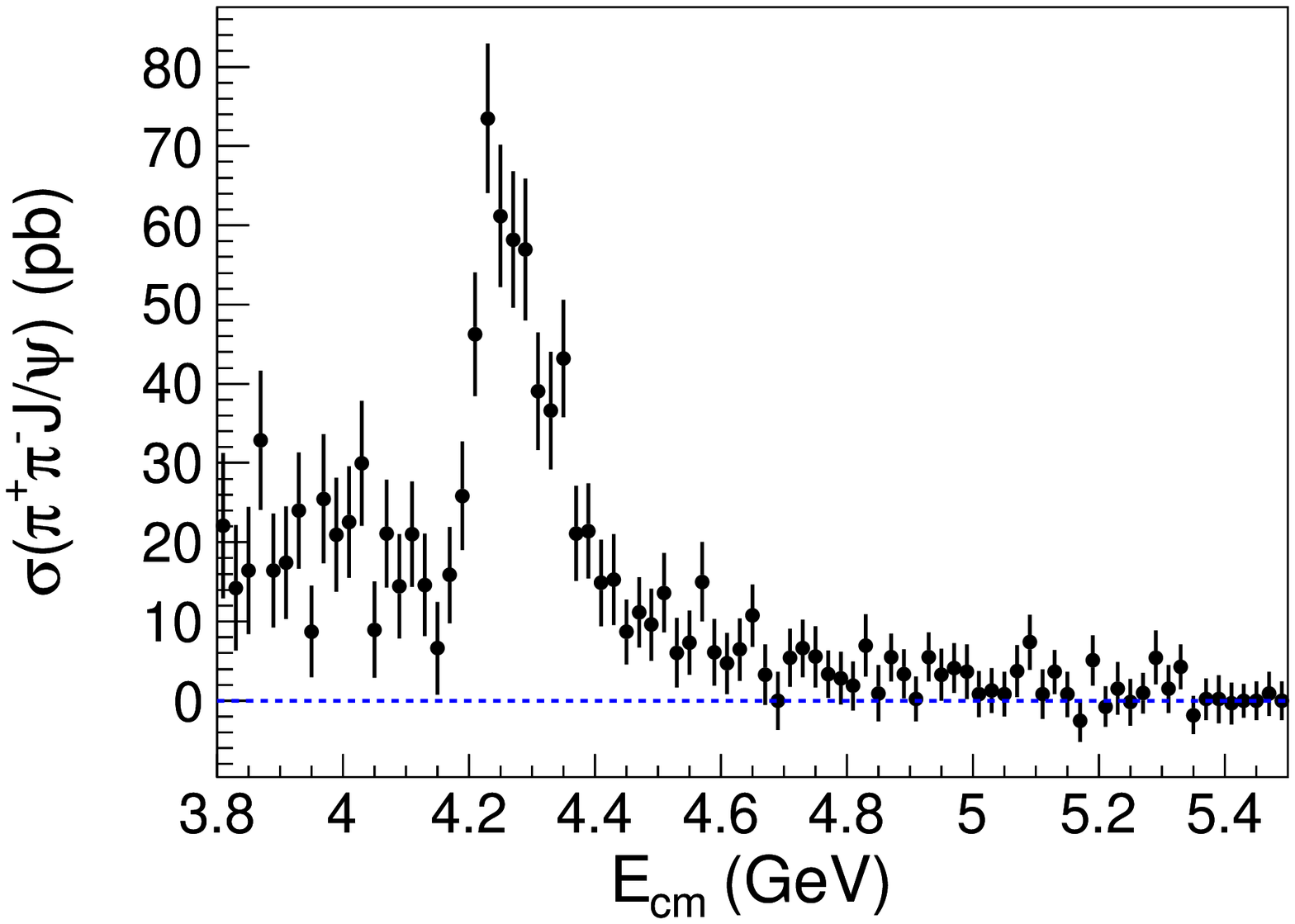}
\caption{(Left) Invariant mass distributions of $\pp\LL$. Points
with error bars are data, and the shaded histograms are the
normalized $\jpsi$ mass sidebands. The solid curves show the total
best fit with two coherent resonances and contribution from
background. The inset shows the distributions on a logarithmic
vertical scale. The large peak around 3.686~GeV/$c^2$ is the
$\psp\to \ppjpsi$ signal. (Right) Cross section  of $\EE\to
\ppjpsi$ after background subtraction. The errors are statistical
only.} \label{mppjpsi}
\end{figure}

The cross section for $e^+ e^- \to \pi^+ \pi^- J/\psi$ between 3.8
and 5.5~GeV is measured with a 967~fb$^{-1}$ data sample collected
by the Belle detector at or near the $\Upsilon(nS)$ ($n = 1,\ 2,\
...,\ 5$) resonances~\cite{belley_new}. Figure~\ref{mppjpsi}(a)
shows the $\pp\LL$ invariant mass distributions after all of these
selection requirements are applied and Fig.~\ref{mppjpsi}(b) shows
the measured cross sections for $\EE \to \ppjpsi$, where the error
bars are statistical only.

\begin{figure}[htb]
 \includegraphics[height=6cm]{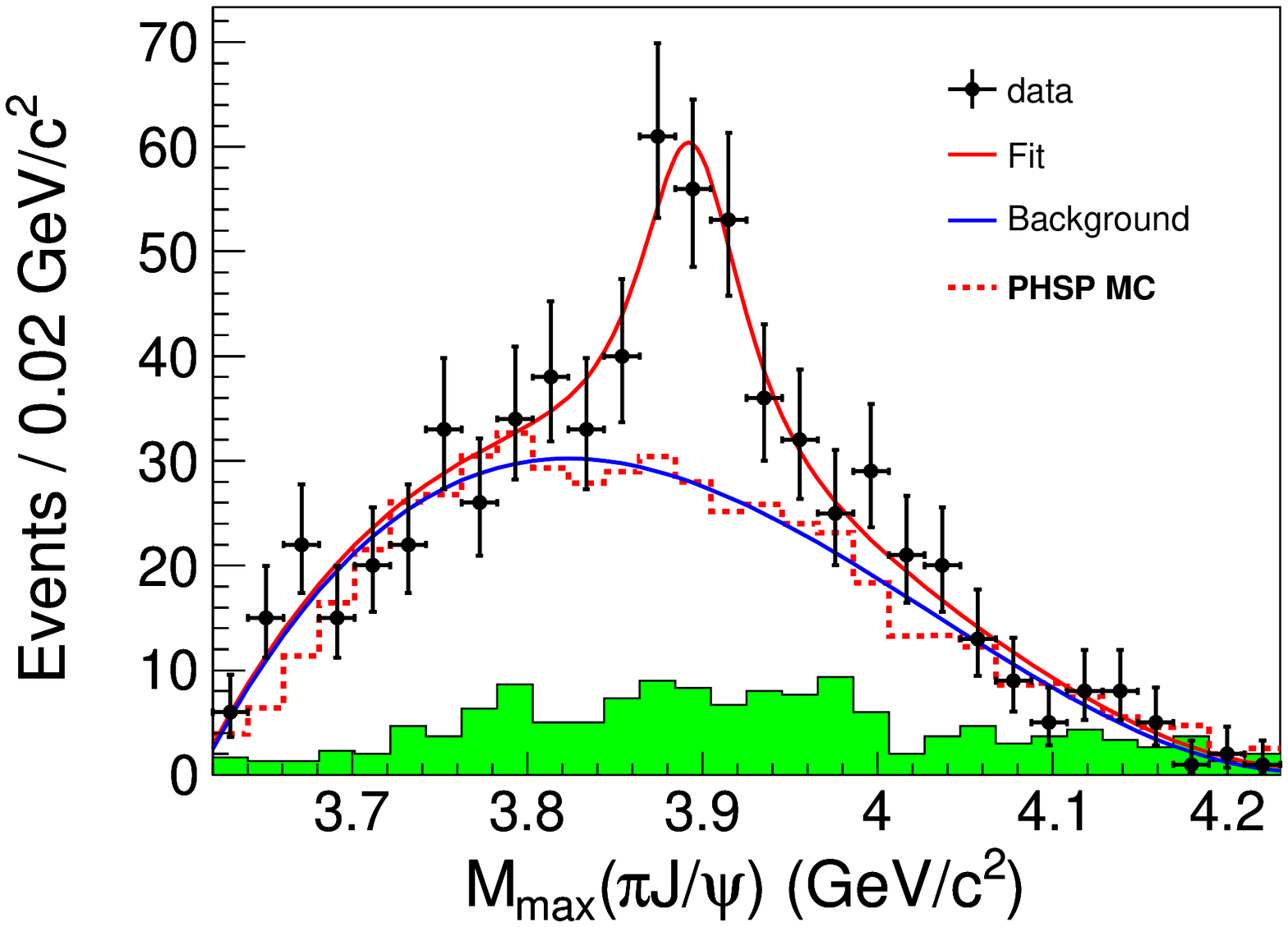}
 \includegraphics[height=6cm]{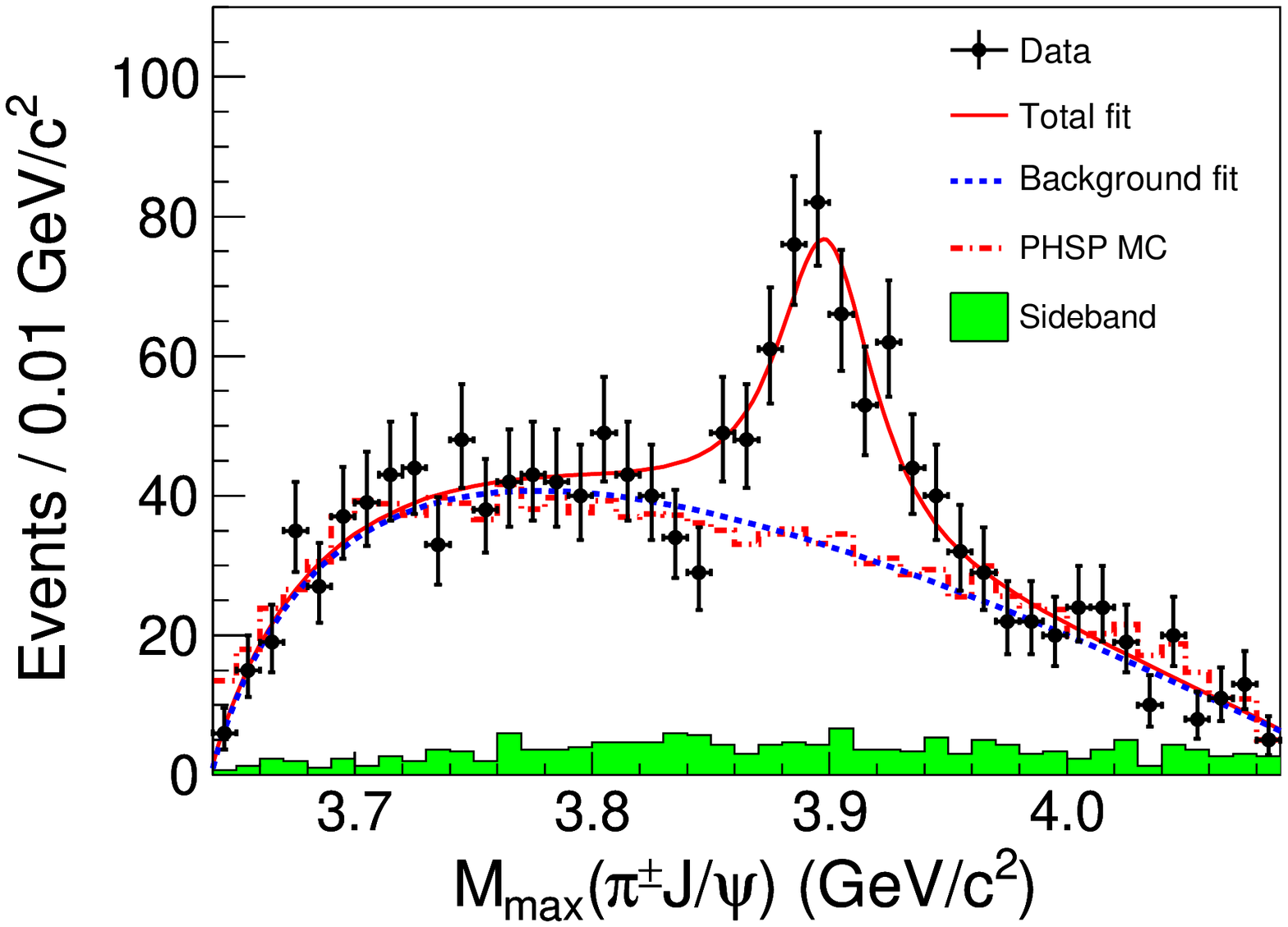}
 \caption{Unbinned maximum likelihood fit to the distribution of
the $M_{\mathrm{max}}(\pi J/\psi)$ (left panel from Belle and
right panel from BESIII). Points with error bars are data, the
curves are the best fit, the dashed histograms are the phase space
distributions and the shaded histograms are the non-$\ppjpsi$
background estimated from the normalized $\jpsi$ sidebands.}
\label{projfit}
\end{figure}

The intermediate states in $\y\to \ppjpsi$ decays are also
investigated~\cite{belley_new}. The $Z(3900)^+$ state (is now
called $\zc$ after BESIII~\cite{zc3900}) with a mass of
$(3894.5\pm 6.6\pm 4.5)~{\rm MeV}/c^2$ and a width of $(63\pm
24\pm 26)$~MeV is observed in the $\pi^\pm\jpsi$ mass spectrum
(see Fig.~\ref{projfit}) with a statistical significance larger
than $5.2\sigma$.

BESIII experiment studied the process $\EE\to \ppjpsi$ at a CM
energy of $4.260$~GeV using a 525~pb$^{-1}$ data
sample~\cite{zc3900}. The $\zc$ is observed in the $\pi^\pm \jpsi$
mass spectrum with a statistical significance larger than
$8\sigma$. A fit to the $\pi^\pm\jpsi$ invariant mass spectrum
(see Fig.~\ref{projfit}), neglecting interference, results in a
mass of $(3899.0\pm 3.6\pm 4.9)~{\rm MeV}/c^2$ and a width of
$(46\pm 10\pm 20)$~MeV. The $\zc$ was confirmed shortly after with
CLEO-c data at a CM energy of 4.17~GeV~\cite{seth_zc}, the mass
and width agree with the BESIII and Belle measurements very well.

\subsection{\boldmath $\EE\to \pppsp$ and $Z_c\to \pi^\pm \psp$}

Using the 980~fb$^{-1}$ full data sample taken with the Belle
detector, the analysis of $\EE\to \pppsp$ is updated with two
$\psp$ decay modes~\cite{belle_pppsp_new}, namely, $\pp\jpsi$ and
$\MM$.

\begin{figure}[htb]
\centering
\includegraphics[width=0.35\textwidth,height=0.45\textwidth,angle=-90]{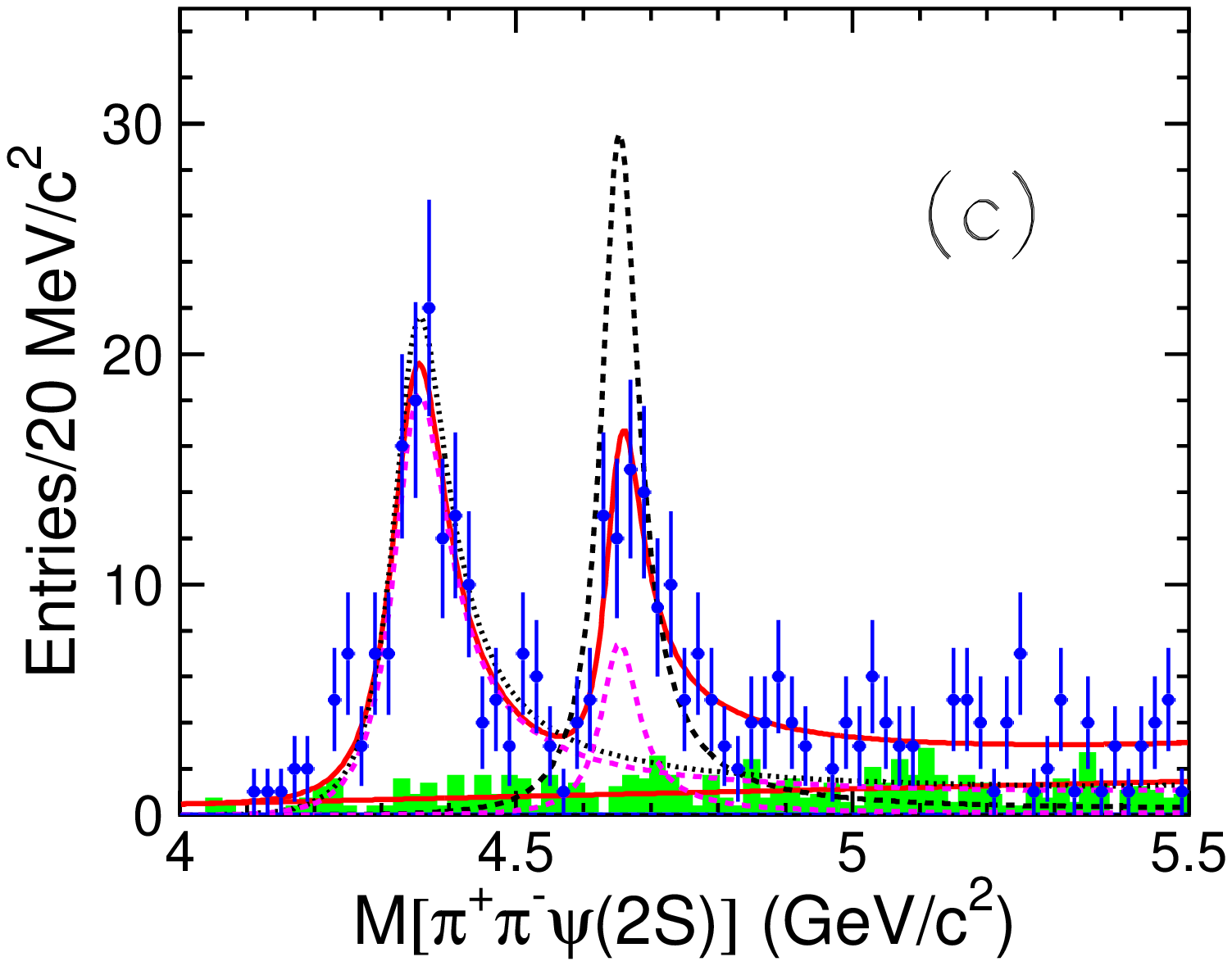}
\includegraphics[width=0.35\textwidth,height=0.45\textwidth,angle=-90]{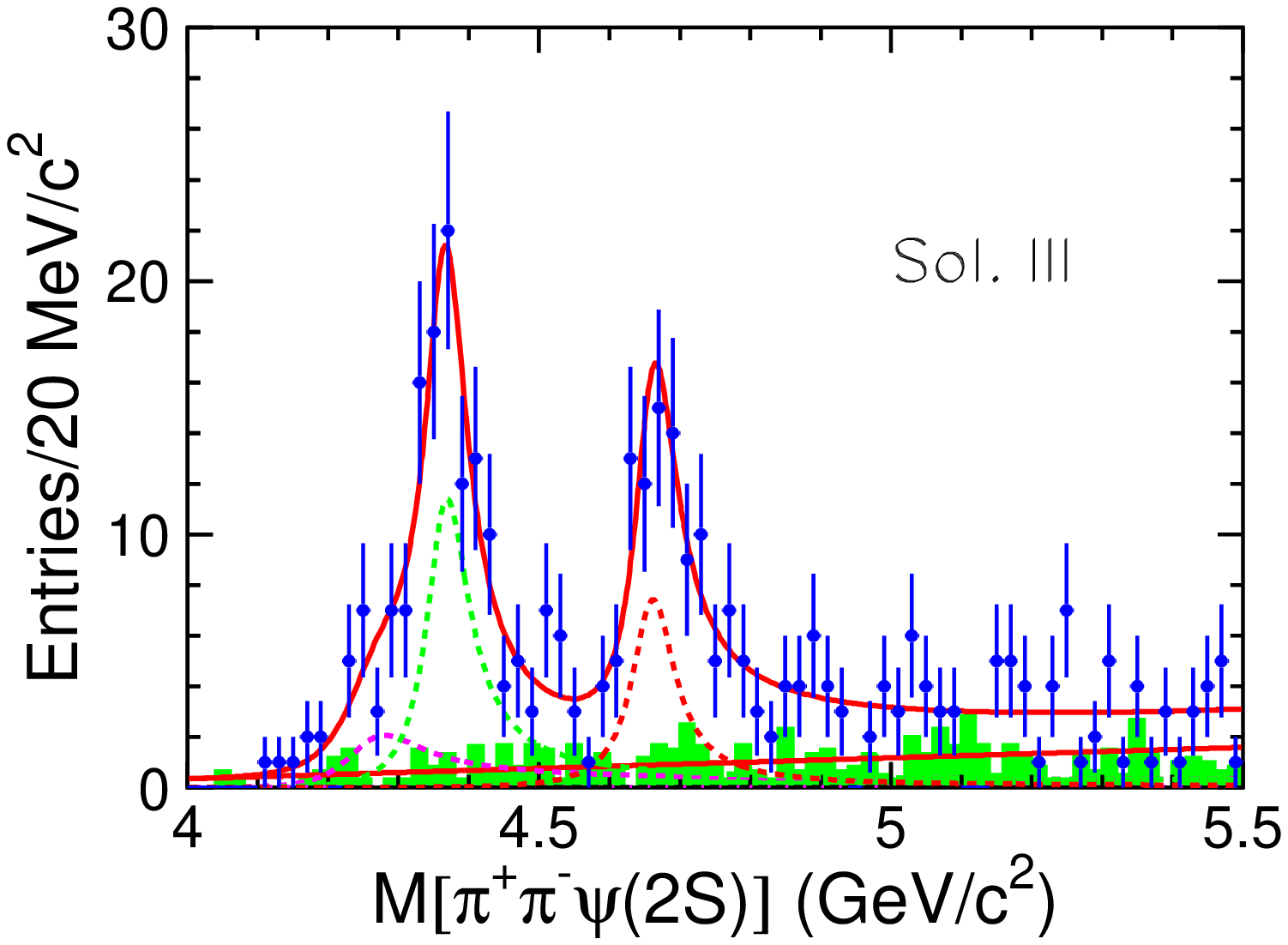}
\caption{The $\pppsp$ invariant mass distribution from the Belle
experiment and the fit results with the coherent sum of two BW
functions (left) and three BW functions (right). The sum of
$\ppjpsi$ and $\MM$ modes is shown. The points with error bars are
data while the shaded histograms the normalized $\psp$ sideband
backgrounds. The curves show the contributions from different BW
components.} \label{2bwfit}
\end{figure}

Fitting the mass spectrum of $\pppsp$ with two coherent BW
functions (see Fig.~\ref{2bwfit}), Belle obtains $M_{Y(4360)} =
(4347\pm 6\pm 3)~\mevcs$, $\Gamma_{Y(4360)} = (103\pm 9\pm
5)~\mev$, $M_{Y(4660)} = (4652\pm10\pm 8)~\mevcs$,
$\Gamma_{Y(4660)} = (68\pm 11\pm 1)~\mev$. and $\BR[Y(4360)\to
\pp\psp]\cdot \Gamma_{Y(4360)}^{\EE} = (10.9\pm 0.6\pm 0.7)~{\rm
ev}$ and $\BR[Y(4660)\to \pp\psp]\cdot \Gamma_{Y(4660)}^{\EE} =
(8.1\pm 1.1\pm 0.5)~{\rm ev}$ for one solution; or $\BR[Y(4360)\to
\pp\psp]\cdot \Gamma_{Y(4360)}^{\EE} = (9.2\pm 0.6\pm 0.6)~{\rm
ev}$ and $\BR[Y(4660)\to \pp\psp]\cdot \Gamma_{Y(4660)}^{\EE} =
(2.0\pm 0.3\pm 0.2)~{\rm ev}$ for the other. Here, the first
errors are statistical and the second systematic.

Since there are some events accumulating at the mass region of
$Y(4260)$, the fit with the $Y(4260)$ included is also performed.
In the fit, the mass and width of the $Y(4260)$ are fixed to the
latest measured values at Belle~\cite{belley_new}. There are four
solutions with equally good fit quality. The signal significance
of the $Y(4260)$ is estimated to be $2.1\sigma$. The fit results
are shown in Fig.~\ref{2bwfit} for one of the solutions. In this
fit, one obtains $M[Y(4360)]=(4365\pm 7\pm 4)$~MeV/$c^2$,
$\Gamma[Y(4360)]=(74\pm 14\pm 4)$~MeV, $M[Y(4660)]=(4660\pm 9\pm
12)$~MeV/$c^2$, and $\Gamma[Y(4660)]=(74\pm 12\pm 4)$~MeV.

\begin{figure}[htb]
\includegraphics[height=0.5\textwidth,angle=-90]{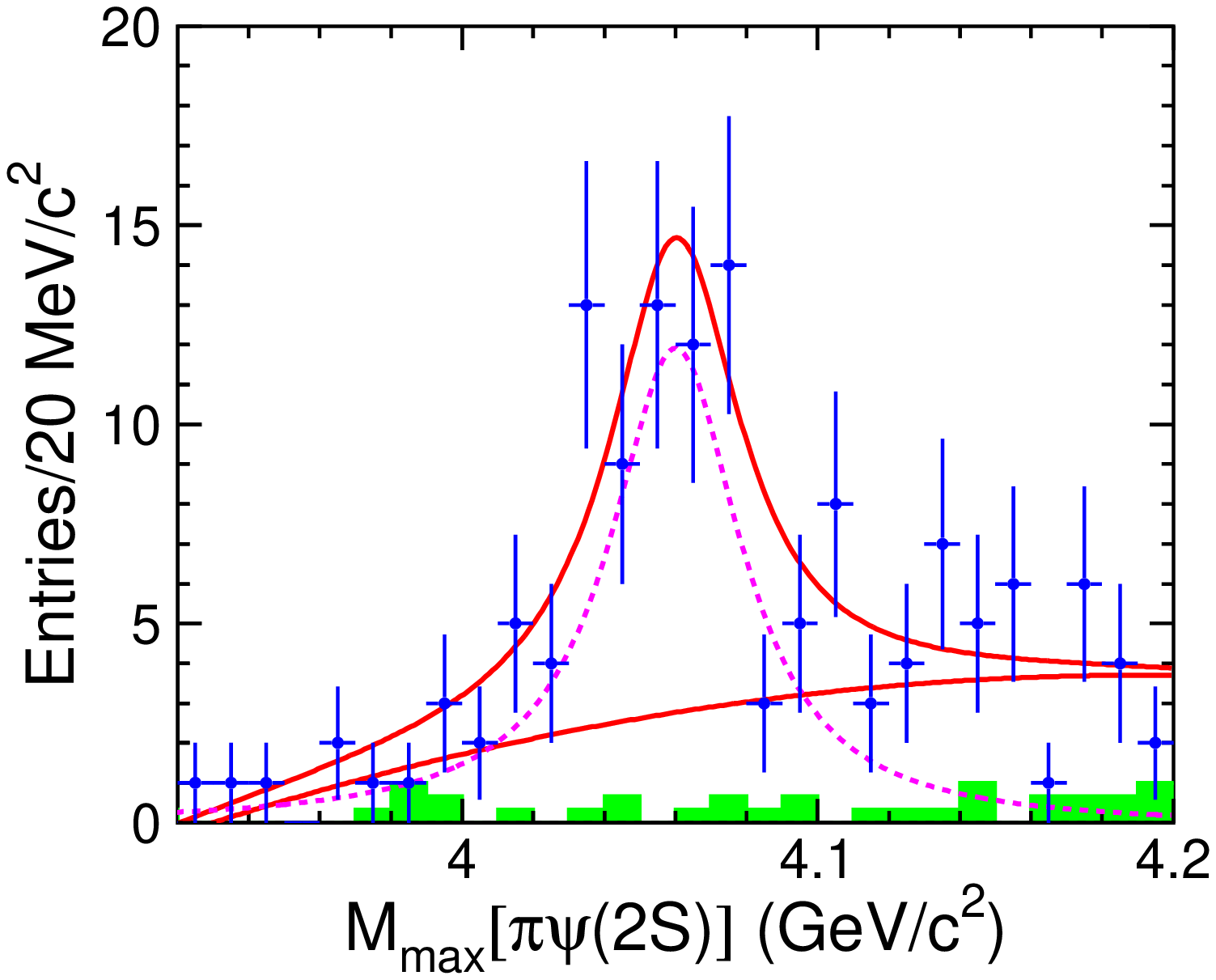}
\caption{The distribution of $M_{\rm max}(\pi^{\pm}\psp)$ from
$Y(4360)$-subsample decays. The points with error bars represent
the data; the histogram is from the sidebands and normalized to
the signal region; the solid curve is the best fit and the dashed
curve is the signal parameterized by a BW function. }
\label{mppsp-fit}
\end{figure}

Possible charged charmoniumlike structures in $\pi^{\pm}\psp$
final states from the $Y(4360)$ or $Y(4660)$ decays are searched
for with the selected candidate events. Figure~\ref{mppsp-fit}
shows $Y(4360)$ decays. An unbinned maximum-likelihood fit is
performed on the distribution of $M_{\rm max}(\pi^{\pm}\psp)$, the
maximum of $M(\pi^+\psp)$ and $M(\pi^-\psp)$, simultaneously with
both the $\ppjpsi$ and the $\MM$ modes. We obtain a mass of
$(4054\pm 3({\rm stat.})\pm 1({\rm syst.}))~\mevcs$ and a width of
$(45\pm 11({\rm stat.})\pm 6({\rm syst.}))~\mev$ for the $Z^\pm$
structure in the $\pi^{\pm}\psp$ system, and the statistical
significance of the signal is $3.5\sigma$. The $Y(4660)$ sample is
very limited in statistics, and there is no significant structures
in the $\pi^\pm\psp$ system.

\subsection{\boldmath $\EE\to \kkjpsi$ and $Z_{cs}\to K^\pm \jpsi$}

\begin{figure}[htb]
 \includegraphics[height=6cm]{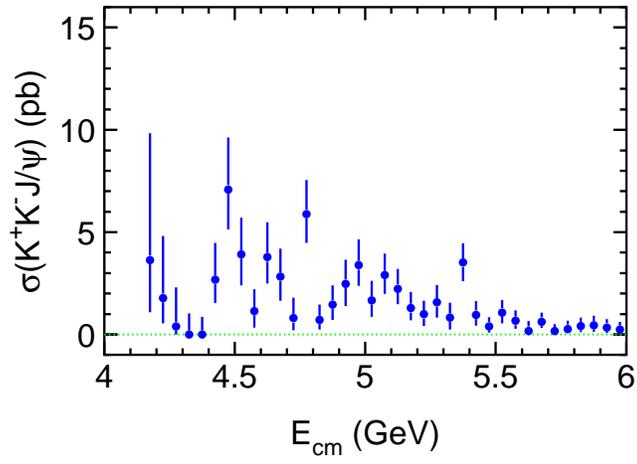}
\caption{The measured $\EE \to \kkjpsi$ cross sections for CM
energies up to 6.0~GeV (points with error bars). The errors are
statistical and a 7.8\% systematic error that is common for all
data points is not included. } \label{cs}
\end{figure}

The cross section of the process $\EE\to \kkjpsi$ are measured via
ISR at CM energies between the threshold and 6.0~GeV using a data
sample of 980~fb$^{-1}$ collected with the Belle detector on or
near the $\Upsilon(nS)$ resonances, where $n=$1, 2, ...,
5~\cite{belle_kkjpsi_new}. The cross sections for $\EE\to \kkjpsi$
are at a few pico-barn level as shown in Fig.~\ref{cs}. Possible
intermediate states for the selected $\kkjpsi$ events are also
investigated by examining the Dalitz plot but no clear structure
is observed in the $K^{\pm}\jpsi$ system. A larger data sample is
necessary to obtain more information about possible structures in
the $\kkjpsi$, $\kk$ and $K^{\pm} \jpsi$ systems.

\section{SUMMARY}

There are lots of progress on the study of the quarkonium and
quarkoniumlike states at Belle. The Belle II
experiment~\cite{belle2} under construction, with about
50~ab$^{-1}$ data accumulated, will surely improve our
understanding of all these states.

\section{ACKNOWLEDGMENTS}

This work is supported in part by National Natural Science
Foundation of China (NSFC) under contract Nos. 11235011 and
11475187; the Ministry of Science and Technology of China under
Contract No. 2015CB856701, and the CAS Center for Excellence in
Particle Physics (CCEPP).

% References

\end{document}